\documentstyle[preprint,eqsecnum,epsfig,prd,aps,floats]{revtex} 

\begin{document}

\newcommand{\drawsquare}[2]{\hbox{%
\rule{#2pt}{#1pt}\hskip-#2pt
\rule{#1pt}{#2pt}\hskip-#1pt
\rule[#1pt]{#1pt}{#2pt}}\rule[#1pt]{#2pt}{#2pt}\hskip-#2pt
\rule{#2pt}{#1pt}}

\newcommand{\Yfund}{\raisebox{-.5pt}{\drawsquare{6.5}{0.4}}}
\newcommand{\Ysymm}{\raisebox{-.5pt}{\drawsquare{6.5}{0.4}}\hskip-0.4pt%
        \raisebox{-.5pt}{\drawsquare{6.5}{0.4}}}
\newcommand{\Yasymm}{\raisebox{-3.5pt}{\drawsquare{6.5}{0.4}}\hskip-6.9pt%
        \raisebox{3pt}{\drawsquare{6.5}{0.4}}}
\newcommand{\Ythree}{\raisebox{-3.5pt}{\drawsquare{6.5}{0.4}}\hskip-6.9pt%
        \raisebox{3pt}{\drawsquare{6.5}{0.4}}\hskip-6.9pt
        \raisebox{9.5pt}{\drawsquare{6.5}{0.4}}}

\newcommand{\nn}{\nonumber}
\newcommand{\eq}[1]{Eq.~(\ref{#1})}
\newcommand{\be}{\begin{equation}}
\newcommand{\ee}{\end{equation}}
\newcommand{\bea}{\begin{eqnarray}}
\newcommand{\eea}{\end{eqnarray}}

\newcommand{\Tr}{{\rm Tr}}
\newcommand{\diag}{{\rm diag}}
\newcommand{\ra}{\rightarrow}
\newcommand{\eps}{\epsilon}
\newcommand{\vareps}{\varepsilon}
\newcommand{\spin}{{Spin(32)/{\bf Z}_2}}
\newcommand{\eeig}{{E_8\times E_8}}

\preprint{\vbox{ \tighten {
		\hbox{MIT-CTP-2882}
		\hbox{PUPT-1882}
		\hbox{IASSNS-HEP-99/63}
                \hbox{hep-th/9907210}
		}  }}  
  
\title{Heterotic Little String Theories and Holography} 
  
\author{ Martin Gremm\thanks{email: gremm@feynman.princeton.edu }\thanks{ 
On leave of absence from MIT, Cambridge, MA 02139}} 
  
\address{Joseph Henry Laboratories, Princeton University, Princeton, NJ 08544} 
 
\author{Anton Kapustin\thanks{email: kapustin@ias.edu}} 
 
\address{ School of Natural Sciences, Institute for Advanced Study\\ 
Olden Lane, Princeton, NJ 08540} 
 
\maketitle  
 
\begin{abstract} 
\tighten{ 
It has been conjectured that Little String Theories in six dimensions are
holographic to critical string theory in a linear dilaton background.
We test this conjecture for theories arising on the worldvolume of heterotic 
fivebranes. We compute the spectrum of chiral primaries in these theories and 
compare with results following from Type I-heterotic duality and the 
AdS/CFT correspondence. 
We also construct holographic duals for heterotic fivebranes near orbifold 
singularities. Finally we find several new Little String Theories which 
have $Spin(32)/{\bf Z}_2$ or $E_8\times E_8$ global symmetry but do 
not have a simple interpretation either in heterotic or M-theory.
} 
\end{abstract} 
 
\newpage

\section{Introduction} 
One of the surprises of the second string revolution was the discovery of
nontrivial Poincare-invariant quantum theories in six 
dimensions~\cite{witten,andy,gh,sw,dlp,dvv,brs,seiberg}.
Some of these theories are superconformal quantum field theories with
$(2,0)$ or $(1,0)$ supersymmetry. Others, dubbed Little String 
Theories (LSTs), are not scale-invariant and do not have a
stress-energy tensor or other local operators. They may have $(2,0), (1,1),$ 
or $(1,0)$ supersymmetry. At length-scales much larger than the nonlocality 
scale they flow either to free super Yang-Mills theories, or to one of the 
nontrivial superconformal field theories mentioned above. 

One way to construct LSTs is to consider flat parallel Neveu-Schwarz
fivebranes in Type II or heterotic string theory and take the limit of 
zero string coupling. The theory on the branes remains interacting
in this limit and defines an LST. Another way is to consider
Type II or Type I string theory on an orbifold of the form 
${\bf R}^4/\Gamma$ for some finite group $\Gamma\subset SU(2)$. In the limit 
of zero string coupling the theory of twisted modes
defines an LST. Finally, one may combine the two methods.

Using these methods the following classes of LSTs have been constructed:

(iia) $(1,1)$ LSTs obtained from IIA on ADE singularities~\cite{seiberg};

(iib) $(2,0)$ LSTs obtained from IIB on ADE singularities~\cite{seiberg};

(o) $(1,0)$ LSTs obtained from $\spin$ heterotic fivebranes in flat space~\cite{seiberg};

(e) $(1,0)$ LSTs obtained from $\eeig$ heterotic fivebranes in flat space~\cite{seiberg};

(ii$'$) $(1,0)$ LSTs obtained from IIA or IIB fivebranes at ADE 
singularities~\cite{intriligator};

(o$'$) $(1,0)$ LSTs obtained from $\spin$ heterotic fivebranes at 
ADE singularities~\cite{intriligator};

(e$'$) $(1,0)$ LSTs obtained from $\eeig$ heterotic fivebranes at
ADE singularities~\cite{intriligator}.

These classes of LSTs are manifestly inequivalent as they have different
supersymmetries and/or global symmetries. Some of these theories have alternative
realizations; for example the LST obtained from IIA (IIB) string theory on an
$A_{N-1}$ 
singularity is equivalent to the LST of $N$ IIB (IIA) fivebranes.
Several other $(1,1)$ LSTs have been constructed in Ref.~\cite{wittennew}.
They can be regarded as ``excitations'' of LSTs of type 
(iia)~\cite{seiberg}.

Above we defined LSTs by considering the decoupling limit of critical string
theories. One can give a more intrinsic definition using Discrete Light Cone
Quantization~\cite{dvv,DLCQ1,DLCQ2,DLCQ3}. Finding a Poincare-invariant
intrinsic definition of LSTs is an important open problem. 

In Ref.~\cite{ABKS} it was conjectured that LSTs have a dual description in terms
of critical string theory in linear dilaton backgrounds. This conjecture is similar
in spirit to the AdS/CFT correspondence~\cite{m} which posits that certain 
conformally-invariant quantum field theories in $d$ dimensions are 
dual to critical string theory or M-theory
on backgrounds of the form $AdS_{d+1}\times X$ for some Einstein manifold $X$.
Just as in the case of the AdS/CFT correspondence, stringy excitations are conjectured
to be in one-to-one correspondence with operators in the boundary theory, and 
their S-matrix is the generating functional for the boundary correlators.
The AdS/CFT correspondence becomes effective in the large-$N$ limit, where 
string/M-theory reduces to supergravity. In contrast, the holographic conjecture
of Ref.~\cite{ABKS} provides a framework in which  the spectrum of operators
in the boundary theory can be calculated for any $N$, but in general 
correlators cannot be computed.
This is because the linear dilaton background by definition includes a strong 
coupling region where string perturbation theory fails. In some cases the
strong-coupling singularity is resolved via M-theory, in which case one can
compute the correlators for large $N$~\cite{ABKS,SM}.

For theories of type 
(iia) and (iib) the relevant linear dilaton background is the familiar ``throat'' 
CFT describing the near-horizon physics of Type II fivebranes. The bosonic fields
of this CFT consist of six free bosons describing coordinates parallel to 
the worldvolume of the branes, a boson describing the radial transverse 
coordinate, and an $SU(2)$ Wess-Zumino-Witten (WZW) model describing the
angular transverse coordinates. The holographic point
of view reduces the classification problem for $(1,1)$ and $(2,0)$ LSTs
to the classification of modular invariant partition functions for
the WZW part of the CFT. It has been argued in Refs.~\cite{ABKS,ds}
that LSTs obtained from Type II on ADE singularities correspond
to ADE modular invariants of $SU(2)$ WZW models~\cite{ciz}. Thus it appears
that (iia)
and (iib) theories represent a complete list of LSTs with sixteen
supercharges. Furthermore, it has been checked in Ref.~\cite{ABKS} that
the spectrum of worldsheet vertex operators in short representations of
the space-time SUSY algebra includes all the expected moduli of the boundary
LST. This supports the holographic conjecture.

In this paper we study heterotic $(1,0)$ LSTs using the holographic approach of
Ref.~\cite{ABKS} (see also \cite{cvj} for related work).
In Section II we construct the holographic description
of heterotic fivebranes in flat space (LSTs of types (o) and (e)). We
compute the spectrum of chiral primaries of the space-time supersymmetry 
algebra and compare with results known from the AdS/CFT correspondence
and heterotic-Type I duality. Our analysis provides evidence
that a single heterotic fivebrane has a nontrivial decoupling limit,
unlike a single fivebrane in Type II. We find that all known operators
of the heterotic LSTs (namely the moduli and their superpartners) have worldsheet
counterparts. 

LSTs of types (o) and (e) have global $SU(2)_L\times SU(2)_R\times 
G$ symmetry, where $G=\spin$ or $\eeig$, respectively. The $SU(2)_R$ factor
is the 
R-symmetry. In Section III we discuss more general $(1,0)$ LSTs with this 
symmetry. For either choice of $G$ the theories we construct are in one-to-one
correspondence with affine Dynkin diagrams of types A,D, or E,
with theories of types (o) and (e) being the A-series. 
As for type II theories, the ADE scheme arises from the ADE classification of
$SU(2)$ WZW modular invariants.
The D-series can be obtained by considering heterotic fivebranes at an $A_1$ 
singularity, i.e. they are a special class of theories of types (o$'$) and 
(e$'$). The exceptional E-series are a new type of LSTs which do not have a 
large $N$ limit.\footnote{We would like to emphasize that the ADE LSTs we are discussing
here are not related to LSTs describing heterotic fivebranes near ADE orbifold 
singularities. The latter theories have smaller global symmetry groups.}
Since the spectrum of chiral operators in the D and E theories
is rather complex, we only discuss a few interesting examples, leaving 
a full investigation for the future.

In general, $(1,0)$ heterotic LSTs need to have only $SU(2)_R$ global symmetry. 
Examples of such theories can be obtained by orbifolding 
ADE heterotic LSTs by a finite group $\Gamma\subset SU(2)_L\times G$.
In Section IV  we consider ${\bf Z}_n$ orbifolds of the A-series which
describe heterotic fivebranes at $A_{n-1}$ singularities. (The holographic
description of Type II fivebranes at $A_{n-1}$ singularities was
discussed in Ref.~\cite{diagom}.) We derive
the constraints following from the modular invariance of the worldsheet.
Interestingly one can construct a consistent worldsheet description of 
an orbifold with an arbitrary embedding of the spin connection into the gauge 
connection.  
This means that the holographic description exists for an arbitrary 
LST of types (o$'$) and (e$'$). In contrast, for toroidal heterotic orbifolds 
(without fivebranes) modular invariance constrains the embedding in such a way
as to ensure that the fivebrane charge of each orbifold plane vanishes.
Section V contains our conclusions.

\section{Holographic description of heterotic fivebranes in flat space}

\subsection{Supergravity solutions}

The solution of the low-energy effective theory describing several
coincident heterotic fivebranes (the ``symmetric fivebrane'') was 
constructed in Ref.~\cite{CHS}. This solution is very similar to the one 
describing Type II fivebranes~\cite{CHS2}, but there are some slight 
differences in the interpretation of parameters, as argued below.

The symmetric fivebrane solution has the form
\bea\label{sugra}
&&A_\mu=-\rho_N^2 \bar\Sigma_{\mu\nu} \frac{2y^\nu}{y^2(y^2+\rho_N^2)},\nn \\ 
&&e^{2\phi}=e^{2\phi_0}+\frac{N\alpha'}{y^2},\nn \\
&&H_{\mu\nu\lambda}=-\eps_{\mu\nu\lambda}^{\ \ \ \rho}\nabla_\rho\phi,\nn \\
&&ds^2=\eta_{ab}dx^a dx^b+e^{2\phi} dy^\mu dy^\mu.
\eea
The notation here is as follows. The Latin indices from the beginning of the
alphabet run from $0$ to $5$ and are raised and lowered with Minkowski metric 
$\eta_{ab}$, the Greek indices run from $0$ to $3$ and are raised and lowered 
with Euclidean metric $\delta_{\mu\nu}$, $y^2$ means $y^\mu y^\mu$, 
$\rho_N^2=N\alpha'\exp(-2\phi_0)$, and
$\bar\Sigma_{\mu\nu}$ is the anti-self-dual t'Hooft tensor taking values
in the Lie algebra of $SU(2)$. It is understood that the $SU(2)$ is
minimally embedded into $\spin$ or $E_8\times E_8$,
i.e. it is identified with one of the $SU(2)$ groups in the
$(Spin(28)\times SU(2)\times SU(2))/{\bf Z}_2$ subgroup of $\spin$ or with the 
$SU(2)$ group in the $E_8\times (E_7\times SU(2))/{\bf Z}_2$ subgroup of 
$\eeig$.

The fivebrane charge of the solution Eq.~(\ref{sugra}) (defined as the integral
of $H$ over an ${\bf S}^3$ $y^2=R^2, R\ra\infty$) is equal to $N$. 
Intuitively, Eq.~(\ref{sugra}) describes a single $SU(2)$ instanton of 
size $\rho_N$ superimposed with $N-1$ small instantons (fivebranes)
located at $y^\mu=0$. This statement requires some qualifications.
Since $dH=0$, the integral of $H$ over $y^2=R^2$ is independent of $R$ and equal 
to $N$. This means that if we define the number of fivebranes at $y=0$ as
the integral of $H$ over a small ${\bf S}^3$ surrounding $y=0$, we get $N$, and
not $N-1$. However, this definition of the fivebrane charge is not unique.
Note that $H$ satisfies a Bianchi identity of the form 
\be\label{Bid}
dH=\Tr\ R(\Omega_-)\wedge R(\Omega_-)-\Tr\ F\wedge F,
\ee 
where $\Omega_-$ differs from the Levi-Civita connection $\omega$ by a torsion
term, $\Omega^{ab}_{-\mu}=\omega_\mu^{ab}-H_\mu^{ab}$. One can define a new
field strength $H_0$ which is a nonlinear function of $H$ and $\omega$
and satisfies the more usual form of the Bianchi identity
\be
dH_0=\Tr\ R(\omega)\wedge R(\omega)-\Tr\ F\wedge F.
\ee
Using $H_0$ instead of $H$ we get a different (gauge-invariant) definition
of the fivebrane charge. For the symmetric fivebrane solution 
the new fivebrane charge measured at infinity is still $N$, while the
fivebrane charge measured on a small ${\bf S}^3$ surrounding $y=0$ is
$N-1$. Given this nonuniqueness of the fivebrane charge, we prefer to
fix the precise relation between $N$ and the number of fivebrane at the origin 
$N_5$ by matching symmetries and operators of the worldsheet CFT and the
boundary LST. We will see below that this leads to the 
identification $N=N_5+1$.\footnote{Note that in the Type II case the relation 
between
$N$ and $N_5$ is simply $N=N_5$.} 

We are interested in the decoupling limit $\phi_0\ra -\infty$ when the
asymptotic string coupling goes to zero. If we introduce the
radial coordinate $t=\log \sqrt{y^2}$ and rescale 
$x^a\ra x^a (N\alpha')^{1/2},$ the limiting solution takes the form
\bea\label{hetlin}
&& A_\mu=-\bar\Sigma_{\mu\nu} \frac{2y^\nu}{y^2},\nn \\ 
&& \phi=-t,\nn \\
&& H=-N\eps,\nn \\
&& ds^2=N\alpha'\left( \eta_{ab} dx^a dx^b + dt^2+ds_3^2\right),
\eea
where $ds_3^2$ is the standard round metric on the hypersurface $y^2=1$
and $\eps$ is its volume form.
Note that the transverse coordinates $y^\mu$ now parametrize a copy of
${\bf S}^3\times {\bf R}$ with $N$ units of $H$-flux on ${\bf S}^3$.
The size of the single instanton goes to infinity in the decoupling limit, 
and the gauge field $A_\mu$ becomes a flat connection on ${\bf S}^3\times 
{\bf R}$. Since ${\bf S}^3\times {\bf R}$ is simply connected, 
$A_\mu$ can be gauged away.
A priori, one could expect this supergravity background to solve the
stringy equations of motion only if the dilaton gradient is small 
($\ll {\alpha'}^{-1/2}$), which requires $N\gg 1$. However, it is well 
known that Eq.~(\ref{hetlin}) corresponds to a soluble worldsheet 
CFT~\cite{CHS} and therefore is an exact solution for all $N$.

It has been proposed in Ref.~\cite{ABKS} that string theory in a 
background of the form
\bea\label{lindil}
&& ds^2=\alpha' \left(\eta_{ab} dx^a dx^b+dt^2+g_{ij} dy^i dy^j\right), \nn \\
&& \phi=-\frac{Qt}{2}
\eea
where $\eta_{ab}$ is the Minkowski metric on ${\bf R}^{d-1,1}$,
is equivalent to a quantum theory without gravity on ${\bf R}^{d-1,1}$.
This equivalence is holographic in the sense that the latter theory
lives at $t=-\infty$ and its correlators can be computed from 
the S-matrix of string excitations in the background Eq.~(\ref{lindil}).
In particular, global symmetries of the ``boundary''
theory correspond to gauge symmetries of the ``bulk'' theory.
Of course, in order to compute the S-matrix string perturbation theory
is not sufficient, as the excitations propagating towards $t=-\infty$
inevitably reach the region of strong coupling. But in order to
compute the \emph{spectrum} of excitations it is sufficient to analyze
the theory near the perturbative region $t=+\infty$. (This analysis 
can miss some purely nonperturbative states~\cite{ABKS}. We will 
see examples of this below.) 

Following the proposal of Ref.~\cite{ABKS}, we conjecture that the
heterotic string background Eq.~(\ref{hetlin}) is
holographic to the worldvolume theory of heterotic fivebranes located
at $t=-\infty$.
The parameter $N$ is related to the number of fivebranes $N_5$ living at $t=-\infty$
via $N=N_5+1$.

\subsection{Worldsheet partition function}

The worldsheet CFT corresponding to the heterotic background 
Eq.~(\ref{hetlin}) has been worked out in Ref.~\cite{CHS}. The bosonic
fields are six free bosons $X^a,a=0,\ldots,5$, an $SU(2)$ WZW model at level
$k=N$, and a Feigin-Fuchs field $t$ with real background charge $\sqrt{2/N}$.
The level of the WZW model is simply the flux of the $H$-field through
${\bf S}^3$. The fermionic fields are ten right-moving fermions and 
32 left-moving fermions. Not all fermions are free: four
of the left-movers and four of the right-movers are coupled
to certain (unequal) pure gauge connections. 
Performing the gauge rotation, it possible to make the fermions free.
This rotation is anomalous and shifts the level of the WZW from $k=N$
to $k=N-2$. Thus we end up with ten free right-moving fermions and 32
free left-moving fermions. The difference between $G=\spin$ and $G=\eeig$
is encoded in the boundary conditions for the left-moving fermions.
It is convenient to bosonize the left-movers into
sixteen free bosons $H_\alpha,\alpha=1,\ldots,16$ living on the weight 
lattice of $G$. 

The $(4,0)$ worldsheet superalgebra combines six of the right-moving fermions
$\psi^a$ with $X^a$, and the remaining right-moving fermions $\psi^\mu$
with $t$ and the WZW field $g(z,\bar z)$. 
The gauged worldsheet supercurrent is right-moving and has the form
\be\label{nonesuper}
G(z)=\eta_{ab}\psi^a\partial X^b+\psi^0\partial\phi+
\sqrt\frac{2}{N}\left[J_{WZW}^i\psi^i+
\psi^1\psi^2\psi^3-\partial\psi^0\right].
\ee 
Here and in what follows the Latin indices from the middle of the alphabet
$i,j,\ldots$ run from $1$ to $3$.
The partition function of this CFT on a torus is 
\be\label{Z}
Z(\tau,\bar\tau)=\frac{1}{2}({\rm Im}\ \tau)^{-2}\eta(\bar\tau)^{-20}\
\eta(\tau)^{-8}
\left(\theta_3(q)^4-\theta_4(q)^4-\theta_2(q)^4)\right)
Z_\phi(\tau,\bar\tau) Z_{WZW}(\tau,\bar\tau) \Theta_G(\bar\tau). 
\ee
Here $Z_\phi$ and $Z_{WZW}$ are the partition functions of the Feigin-Fuchs 
field and the $SU(2)$ WZW model, respectively, and $\Theta_G$ is the
theta-function of the (even, self-dual) weight lattice of $G$. It is easy to
check that $Z(\tau,\bar\tau)$ is modular-invariant provided 
$Z_{WZW}(\tau,\bar\tau)$ is a modular-invariant combination of affine 
$SU(2)$ characters. 

A celebrated theorem~\cite{ciz} states that such modular-invariant combinations
have an ADE classification. Only the A-type modular invariant
exists for all levels $k$, and thus it is very plausible that it corresponds
to heterotic fivebranes in flat space. 
We will discuss the interpretation of D and E invariants in the next section.

\subsection{Symmetries of the worldsheet CFT}
Let us now discuss the symmetries of the worldsheet CFT and the corresponding
LSTs. The internal CFT has $(4,0)$ worldsheet
supersymmetry. This implies~\cite{FMS} that the space-time theory has
$(1,0)$ supersymmetry in six dimensions. The corresponding $SU(2)$ R-current
is right-moving and has the form
\be
J_R^i(z)=J_{WZW}^i(z)+\frac{1}{2}\vareps^{ijk}\psi^j\psi^k, \qquad i=1,2,3.
\ee
Its level is $k+2$.\footnote{The right-moving current $J_{WZW}(z)$ is not 
BRST-invariant.}
The supercharges transform in the $({\bf 4},{\bf 2})$ of 
$Spin(1,5)\times SU(2)_R$, where $Spin(1,5)$ is the six-dimensional 
Lorenz group. 
There is also a left-moving $SU(2)_L$ current $J_{WZW}(\bar z)$ with level $k$
and a left-moving $G$ current $J_G(\bar z)$ with level $1$ obtained from the
$H_\alpha$
using the Frenkel-Kac-Segal construction. The supercharges are inert under 
$SU(2)_L\times G$.

The construction of the supercharges is standard~\cite{FMS}.
Bosonizing right-moving
fermions $\psi^a$ and $\psi^\mu$ we can construct spin operators
$S_\alpha$ and $\Sigma_A$, where $\alpha$ is the index of the ${\bf 4}$
representation of $Spin(1,5)$ and $A$ is the index of the ${\bf 2}$ 
representation of $SU(2)_R$. Let $\sigma$ be the bosonized superconformal
ghost. In the $-1/2$ picture the supercurrent is given by 
\be
Q_{\alpha A}(z)=e^{-\sigma(z)/2}S_\alpha(z)\Sigma_A(z).
\ee
The GSO projection is imposed by requiring that physical vertex operators
be local with respect to $Q_{\alpha A}(z)$.

By the usual ``holographic'' logic, the boundary LST 
must have $(1,0)$ SUSY and $SU(2)_L\times SU(2)_R\times G$ global symmetry.
The $SU(2)_R$ symmetry is the R-symmetry of the $(1,0)$ theory.
This is the expected result for a theory of heterotic fivebranes in
flat space. The symmetry $SO(4)\simeq SU(2)_L\times SU(2)_R$
corresponds to the rotations of the directions transverse to the fivebranes.

The case $k=0$ is somewhat special in that the WZW part is absent. Thus the
$SU(2)_L$ symmetry acts trivially in this case. This fact together
with Type I-heterotic duality provides a check
on our identification $k=N-2=N_5-1$. For $G=\spin$ we can think of $N_5$ 
heterotic fivebranes as $N_5$ Type I D5 branes. (The Type I description is not 
suitable for discussing the decoupling limit, but it is good enough to identify 
symmetries and flat directions.) The worldvolume theory of $N_5$ D5 fivebrane 
reduces at low energies to super Yang-Mills theory with gauge group $Sp(N_5)$,
32 hypermultiplets in the fundamental representation, and a hypermultiplet
in the antisymmetric tensor representation. The antisymmetric tensor is
a doublet of $SU(2)_L$, while the rest of the fields are singlets.
For $N_5=1$ the antisymmetric tensor is decoupled from the rest of the theory
(it describes the center-of-mass motion of the D5 brane), and $SU(2)_L$
acts trivially on the low-energy degrees of freedom. According to our
identification, precisely at this value of $N_5$, the $SU(2)_L$ action on
the worldsheet becomes trivial, so apparently $SU(2)_L$ acts trivially on
the full LST, and not just on the low-energy degrees of freedom. 

\subsection{The spectrum of space-time chiral primaries}

The holographic worldsheet description of heterotic LSTs found above can be 
used to compute the spectrum of operators in chiral multiplets of the 
space-time SUSY algebra. For $G=\spin$ we will compare our results
to the spectrum of chiral operators in the super Yang-Mills theory living on
Type I D5 branes. For $G=E_8\times E_8$ we will compare to the results
obtained using the AdS/CFT correspondence~\cite{m,berkooz,gimpop}.

The space-time chiral operators look very much like in the Type II
LSTs~\cite{ABKS}.
For simplicity, we will limit ourselves to space-time scalars.
Consider the following ansatz for the ``internal'' part of the Neveu-Schwarz 
vertex operators (in the $-1$ picture):
\be\label{physop}
V_{j_L,j_R}(z,\bar z)\sim e^{-\sigma(z)} W_{j_L}(\bar z) \psi^i(z) V_{j_R}(z) 
e^{\beta\phi(z,\bar z)}.
\ee
Here $V_{j_R}(z)$ is a primary of the right-moving bosonic $SU(2)$ current 
algebra with $SU(2)$ spin $j_R$, and $W_{j_L}(\bar z)$ is a primary
of the left-moving Virasoro algebra with $SU(2)_L$ spin $j_L$. We need 
to require that this operator be a primary of the gauged supercurrent $G(z)$.
One can show that this condition is met if the $SU(2)_R$ indices of the fermion 
$\psi^i$ and of $V_{j_R}$ are contracted to form the representation ${\bf 2j_R+3}$
or ${\bf 2j_R-1}$. In order for $V_{j_L,j_R}$ to be a vertex operator for
a physical state, it must have dimension $(1,1)$. This is achieved
by setting $\beta=j_R\sqrt{2/N}$ and taking $W_{j_L}(\bar z)$ to be a 
Virasoro primary of dimension
\be\label{dim}
h_L=1+\frac{j_R(j_R+1)}{N}.
\ee
It is easy to check that $V_{j_L,j_R}$ is local
with respect to $Q_{\alpha A}$. Moreover, one can check that it is 
space-time chiral, i.e. acting on it with a space-time supercharge does not
produce operators with higher $SU(2)_R$ spin. Moreover, out of the
two ``physical'' contractions of $SU(2)_R$ indices of $\psi^i$ and
$V_{j_R}$ only the one with $SU(2)_R$ spin $j_R+1$ is a primary (the other 
one is its descendant). Consequently, for any $j_L$ and any acceptable 
$W_{j_L}(\bar z)$ we get precisely one chiral primary of the space-time 
SUSY algebra.

The $A_{N-1}$ modular invariant at level $k=N-2$ contains affine $SU(2)_k$
primaries of 
the form $({\bf 2j+1},{\bf 2j+1})$ with $2j=0,1,\ldots,N-2$. 
Then Eq.~(\ref{dim}) suggests a natural way to get a suitable $W_{j_L}$: one
has to take a left-moving primary $V_{j_R}(\bar z)$ and get its level-one
descendant with respect to the left-moving current algebra.
The full left-moving current algebra is $SU(2)_L\times {\bf R}\times G$,
where the ${\bf R}$-current is $\partial\phi$. 

If $W_{j_L}$ is the descendant with respect to the $G$ current algebra, then
$j_L=j_R$ and $W_{j_L}$ is automatically a Virasoro primary. In this case the 
operator Eq.~(\ref{physop}) transforms as $({\bf 2j+1},{\bf 2j+3})$ of 
$SU(2)_L\times SU(2)_R$ and is in the adjoint of $G$.

If $W$ is a descendant with respect to the $SU(2)_L\times {\bf R}$ current algebra,
then depending on how $SU(2)_L$ indices are contracted we get one operator
with $j_L=j_R+1$, one with $j_L=j_R-1$, and two with $j_L=j_R$. The
first two are automatically physical (i.e. the left-moving part is a Virasoro
primary). One linear combination of the other two is also physical.
Thus we get three physical operators, 
$({\bf 2j_R+3},{\bf 2j_R+3})$, $({\bf 2j_R+1},{\bf 2j_R+3})$, 
and $({\bf 2j_R-1},{\bf 2j_R+3})$. All are singlets with respect to $G$.

To summarize, the theory has $G$-singlet chiral primaries 
of the form $({\bf 2j+3-2n},{\bf 2j+3})$ with $2j=0,1,\ldots,N-2$, $n=0,1,2$,
and chiral primaries in the adjoint of
$G$ of the form $({\bf 2j+1},{\bf 2j+3})$ with $2j=0,\ldots,N-2$. 
Note that $({\bf 2j+3},{\bf 2j+3})$ is a traceless symmetric tensor of
$SO(4)\simeq SU(2)_L\times SU(2)_R$ with $2j+2$ indices. These operators
are in one-to-one correspondence with Casimirs of $SU(N)$. 

{\it Comparison with Type I D5 branes}

Let us now compare these operators to the chiral operators of the super 
Yang-Mills theory on $k+1 = N-1 = N_5$ Type I D5 branes (for $G=\spin$).
The fivebrane
gauge group is $Sp(k+1)$, and the hypermultiplet content is given by
\bea\label{hypers}
&& q\sim ({\bf 1, 2, 32, 2k+2}), \quad Y\sim ({\bf 2, 2, 1, (k+1)(2k+1)-1}) 
\nn \\
&& {\rm of}\ SU(2)_L\times SU(2)_R\times\spin\times Sp(k+1).
\eea
$Y$ is the antisymmetric ``traceless'' tensor of $Sp(k+1)$.

Let us begin with the case $k=0$. In this case the only gauge-invariant
chiral primary operator is $qq$ where the $SU(2)_R$
indices are symmetrized and $\spin$ indices are antisymmetrized.
This operator transforms as ${\bf 3}$ of $SU(2)_R$ and as the adjoint
of $\spin$.\footnote{Incidentally, this operator is the lowest component of a 
supermultiplet containing the $\spin$ global current.} This agrees precisely with
what we found from the worldsheet. To see this note that for $k=0$ the left-moving 
$SU(2)_L$ current is absent, and most of the worldsheet operators listed above
are absent too. 

For $k>0$ the Yang-Mills theory has chiral primary operators in the adjoint of $\spin$
of the form $q Y^p q, p=0,\ldots, k$. They transform as $({\bf p+1,p+3})$
of $SU(2)_L\times SU(2)_R$ and match precisely all the operators
in the adjoint of $\spin$ that we found on the worldsheet. The Yang-Mills spectrum
also includes singlets of $\spin$ of the form $\Tr\ Y^{p+2}, p=0,\ldots,k-1$
which transform as $({\bf p+3, p+3})$ of $SU(2)_L\times SU(2)_R$. They
also have worldsheet counterparts. However, there are other operators
in the LST which do not have counterparts in Yang-Mills theory.
Firstly, there are singlets of $\spin$ with unequal $SU(2)_L$ and
$SU(2)_R$ spins. Secondly, there is a $\spin$-singlet which transforms
as $({\bf k+3, k+3})$ of $SU(2)_L\times SU(2)_R$. 

The latter operator is
interesting. It has the same worldsheet origin as the operators which
match $\Tr\  Y^{p+2}, p=0,\ldots,k-1$ and its quantum numbers are the same
as those of $\Tr\  Y^{p+2}$ for $p=k$. Why is this operator absent in the super
Yang-Mills spectrum? We see three possible explanations for this.

1. In general the operators of the low-energy Yang-Mills theory are a subset of
the operators of the LST. Thus it is not surprising that there is no
one-to-one match between them. In fact, examples of this sort already appear
for LSTs with sixteen supercharges studied in Ref.~\cite{ABKS}. If this interpretation
is correct, then $({\bf k+3, k+3})$ does not correspond to a flat direction of the
LST, unlike $({\bf p+3,p+3}), p=0,\ldots,k-1$. Unfortunately we do not know how
to check this using worldsheet methods.

2. Recall that strictly speaking the symmetric fivebrane solution describes $k+2$
fivebranes with one fivebrane blown up to a finite size $\rho_N$. $\rho_N$
goes to infinity in the decoupling limit, which means that we are infinitely far along the 
Higgs branch in the direction where the gauge group $Sp(k+2)$ is broken down to
$Sp(k+1)$. The effective low-energy Yang-Mills theory includes all the
fields of the $Sp(k+1)$ system plus a decoupled singlet which describes the relative
position of the $k+1$ small fivebranes and the remaining large fivebrane.
We could incorporate this by working with a field $\tilde Y$ which is an antisymmetric
tensor of $Sp(k+2)$, in which case the independent operators are 
$\Tr\ {\tilde Y}^{p+2}, p=0,\ldots,k.$ Naively, we would expect these operators to be expressible as
polynomials in $\Tr\  Y^p$ and the decoupled singlet. The fact that this factorization
does not occur on the worldsheet can be interpreted as the failure of decoupling in
the full LST. If this is correct, then the LST we are studying is \emph{not} the
same as the LST of $k+1$ fivebranes at the origin of their moduli space. 

3. The holographic conjecture is wrong.

In our view, the third possibility is unlikely. To decide between the first
two possibilities one has to establish whether there is a flat directions
along which $({\bf k+3, k+3})$ has a vacuum expectation value.

{\it Comparison with AdS/CFT}

If we take $G=E_8\times E_8$, then the corresponding LST is believed to flow
in the infrared to an interacting $(1,0)$ superconformal field theory. 
Its large $N$ limit is described by 
supergravity on $AdS_7\times {\bf S}^4/{\bf Z}_2$~\cite{berkooz}. To describe the 
${\bf Z}_2$ action
it is best to think of ${\bf S}^4$ as a hypersurface $x_1^2+x_2^2+x_3^2+x_4^2+x_5^2=1$
in ${\bf R}^5$; then ${\bf Z}_2$ acts by $x_5\ra -x_5$ and by flipping the sign of
the M-theory 3-form. The fixed set of the ${\bf Z}_2$ action is an 
${\bf S}^3\subset{\bf S}^4$. Kaluza-Klein excitations on $AdS_7$ are believed to be 
in one-to-one correspondence with operators of the large $N$ $(1,0)$ field theory on
the boundary of $AdS_7$. The spectrum of Kaluza-Klein excitations
has been computed in~\cite{gimpop}. To make a meaningful comparison,
it is useful to keep in mind that although every chiral primary of the 
superconformal algebra at large $N$ is expected to correspond to a chiral operator in 
the LST, the converse is not necessarily true. Some chiral operators in the LST
may acquire 
large (i.e. divergent in the large $N$ limit) dimensions as one flows to the infrared 
and will not be present in the Kaluza-Klein spectrum. 

One can easily see that all chiral primaries charged under $E_8\times E_8$ found in 
Ref.~\cite{gimpop} are reproduced by our worldsheet computation. 
The traceless symmetric tensors of $SO(4)\simeq SU(2)_L\times SU(2)_R$ with
orders $2,3,\ldots$ also appear in the large $N$ analysis. Of course, the
truncation of the spectrum at order $N$ cannot be seen at the level of supergravity.
However, most chiral primaries found in Ref.~\cite{gimpop} (those
which contain odd place-holder fields; see Ref.~\cite{gimpop} for details) do not seem 
to have a ``worldsheet'' counterpart.
After a little thought, this may not be so surprising. The existence of
operators with odd place-holders is due to the presence of the hidden 
eleventh dimension in the $E_8\times E_8$ heterotic string at strong coupling.
One cannot expect to see this hidden dimension by doing a worldsheet CFT computation.
In particular, all operators which are self-dual 3-forms are missed in the
worldsheet approach. Similar phenomena are observed in $(2,0)$ LSTs~\cite{ABKS}.
In the Type II case, however, it may be  possible to find the missing states by considering
D0-branes of the throat CFT~\cite{ABKS}. This has no analogue in the heterotic case, since
the heterotic ``D0-charge'' (the momentum along the eleventh dimensions) is not
conserved.

Similarly, for $G=\spin$ we do not see any evidence for $Sp(k+1)$ gauge symmetry
in our worldsheet approach, because the gauge group arises through
nonperturbative effects (see also \cite{cvj}). The spectrum of chiral
operators is nevertheless consistent with 
heterotic-Type I duality, as discussed above.

\section{More general heterotic LSTs with $SU(2)_L\times SU(2)_R\times
G$ symmetry}

\subsection{Worldsheet partition functions}

In the previous section we discussed the worldsheet description of heterotic 
fivebranes for which the partition function of the WZW model was of the A-type. 
However, we can replace the WZW partition function in \eq{Z} by any other 
modular invariant WZW partition function. Such partition functions have an ADE
classification \cite{ciz}, so we can replace the A-type partition function in
\eq{Z} by a partition function of D- or E-type. The total partition function will
then be modular invariant and we obtain additional models with 
$SU(2)_L\times SU(2)_R\times G$ global symmetry and $(4,0)$ worldsheet
supersymmetry. The A-type partition functions exist for all
levels $k$, the D-type modular invariants require even $k$, and the E-type
partition functions, $E_6,E_7,E_8$, exist for $k = 10,16,28$. 

The models with D-type WZW partition functions have a simple space-time
interpretation. 
Recall that the D-type modular invariants can be regarded
as $Z_2$ orbifolds of the A-type modular invariants at the same level.
The $Z_2$ in question acts on left-moving $SU(2)$ primaries as 
the center of $SU(2)$. In particular it acts on the group-valued
field $g(z,\bar z)$ of the WZW model as
\be
g(z,\bar z)\ra -g(z,\bar z)
\ee
Consequently this orbifolding corresponds to reflecting all four coordinates
transverse to the fivebranes. Thus the D-series describe fivebranes
at an ${\bf R}^4/{\bf Z}_2$ singularity, i.e. they correspond to a particular 
class of theories of types (o$'$) and (e$'$). 
Note that the ${\bf Z}_2$ orbifold action 
here does not involve the internal left-moving bosons, which implies that the gauge 
connection has a trivial monodromy.\footnote{The D-series heterotic CFT we are
discussing can be obtained from the Type II CFT describing fivebranes near
an ${\bf R}^4/{\bf Z}_2 (-1)^{F_L}$ orbifold~\cite{ABKS,diagom} by embedding the spin 
connection into the gauge connection~\cite{gepner}. When applied to the Type II
CFT describing fivebranes near an ${\bf R}^4/{\bf Z}_2$ orbifold, this procedure
yields a slightly different heterotic CFT discussed at the end of Section IV. Its
symmetry group coincides with $SU(2)_L\times SU(2)_R\times G$ locally, 
but not globally.}
By virtue of the anomalous Bianchi identity $dH=\Tr\ R\wedge R - \Tr\ F\wedge F$
such an orbifold carries nonzero fivebrane charge.
The usual perturbative ${\bf R}^4/{\bf Z}_2$ orbifold, on the other hand, 
has the spin connection embedded in the gauge connection in a particular manner
so that the fivebrane charge cancels between $\Tr\ R\wedge R$ and $\Tr\ F\wedge F$
and the group $G$ is broken to a subgroup.
We will discuss fivebranes near this and other orbifolds in the next section.

We still need to determine the precise relation between the number of fivebranes 
$N_5$ and the WZW level $k$. Analogy with the previous section suggests that $k$ is the 
total fivebrane charge counted on the cover up to a shift, i.e. $k=2(N_5-a)$ for some 
constant $a$. 
For $G=\spin$ we can try to determine $a$ by examining the Type I dual of our brane 
configuration. It consists of $N_5$ Type I D5 branes
near an ${\bf R}^4/{\bf Z}_2$ singularity, with ${\bf Z}_2$ acting trivially
on the gauge bundle. Unlike in the case of Type I fivebranes in flat space, at
the origin of the moduli space the fivebrane theory flows to an 
interacting $(1,0)$ fixed point \cite{aspinwall,intriligator2}. 
This theory has a Higgs  branch and, for $N_5\ge 4$, also a Coulomb branch. 
The physics of the Higgs branch is ``exotic'', i.e. it cannot be described by a 
weakly coupled field theory. The physics of the Coulomb branch is described
by super Yang-Mills theory coupled to a tensor multiplet.  
The Yang-Mills theory has gauge group $Sp(N_5)\times Sp(N_5-4)$, 16 
hypermultiplets $q$ in $({\bf 2N_5},{\bf 1})$, and one hypermultiplet $Q$ in 
$({\bf 2N_5},{\bf 2N_5-8})$~\cite{aspinwall,intriligator2}.
The global symmetry of this theory is $SU(2)_L\times SU(2)_R \times \spin$,
where the lowest components of $q$ transform as $({\bf 1,2,32})$ and the lowest
components of $Q$ transform as $({\bf 2,2,1})$. For $N_5=4$ 
the Yang-Mills theory reduces to $Sp(N_5)$ with 16 hypermultiplets
in the fundamental representation, and therefore the
$SU(2)_L$ symmetry acts trivially
on the Coulomb branch. In the worldsheet theory $SU(2)_L$ disappears for $k=0$,
while for $k>0$ the $SU(2)_L$ symmetry clearly acts nontrivially.
This suggests that $k=2(N_5-4)$. Although it is plausible that for $G=\eeig$
the answer is the same, we do not know how to show this.

The theories with E-type WZW partition functions do not have a simple geometric
origin. It is easy to see that they cannot be obtained by orbifolding the 
heterotic fivebrane theory: any orbifold except the ${\bf R}^4/{\bf Z}_2$ that
gives rise to the D-series would break the $SU(2)_L\times G$ symmetry.
On the other hand, the E-series LSTs have the full 
$SU(2)_L\times SU(2)_R\times G$ symmetry group. This implies that the E-series,
unlike the D-series, cannot be obtained as a geometric orbifold of the A-series.
Thus the E-series are new heterotic LSTs.
Since they arise at $k=10,16,28$, and the level is related
to the flux of the H-field through the 3-sphere surrounding the fivebranes
via $k=N-2$, these theories describe systems of $12,18,$ and $30$ fivebranes,
respectively. However, it is not possible to think of these fivebrane
configurations in terms of low-energy supergravity, since the dilaton gradient
(=background charge of the Feigin-Fuchs field) is of order one in string units.

We will not perform an exhaustive analysis of the chiral primary operators of
the D and E models. Instead we discuss a few examples in the next subsection.
For large $k$ it should also be possible to study the D-series LSTs with
$G=\eeig$ using the AdS/CFT correspondence, but this has not been done yet.
The E-series theories exist only for small values of $k$, so the worldsheet 
approach is the only possibility in this case.

\subsection{Examples of chiral primaries}

{\it The D-series}

The modular invariant of type $D_{N/2+1}$ with $N$ even contains ``untwisted''
primaries of the form $({\bf 2j+1},{\bf 2j+1})$ with $2j=0,2,4,\ldots,N-2$
and ``twisted'' primaries of the form $({\bf N/2+2n},{\bf N/2-2n})$
with integer $n$. The left-right symmetric primaries are the easiest to analyze,
as they behave in the same way as the primaries of the A-series. 
The ``untwisted'' primaries are all left-right symmetric and produce the
same spectrum of chiral operators as the A-series, except that $2j$ is restricted
to be even. In particular, we obtain chiral primaries which are traceless symmetric
tensors of $SO(4)$ of even orders $2,4,\ldots,N$. They correspond to the Casimirs
of $SO(N+2)$ of the form $\Tr\ X^{2k}, k=1,\ldots,N/2$. Among the ``twisted''
primaries there is a single left-right symmetric one,
$({\bf N/2},{\bf N/2})$. It yields a chiral primary operator which
is a traceless symmetric tensor of $SO(4)$ of order $N/2+1$. Its order matches
the order of the Pfaffian of $SO(N+2)$. It also yields a chiral primary
$({\bf N/2},{\bf N/2+2})$ which transforms in the adjoint of $G$.

We will not describe the complete spectrum of chiral operators arising
from the left-right asymmetric WZW primaries. Instead, let us consider
a simple example, that of $({\bf N/2+2},{\bf N/2-2})$. In this case the condition
Eq.~(\ref{dim}) is satisfied only if we take $W_{j_L}(\bar z)$ to be the 
primary of the left-moving bosonic $SU(2)$ with spin $(N+2)/4$. Then we
get a space-time chiral primary with $SU(2)_L\times SU(2)_R$ content 
$({\bf N/2+2},{\bf N/2})$.

For $G=E_8\times E_8$ the theory of $N_5$ heterotic fivebranes at an 
${\bf R}^4/{\bf Z}_2$ singularity flows to an interacting $(1,0)$ fixed 
point~\cite{intriligator}. At large $N_5$ this fixed point is expected to be 
holographic to supergravity on $AdS_4\times {\bf S}^4/({\bf Z}_2\times {\bf Z}_2)$,
where the two ${\bf Z}_2$'s act by $x_5\ra -x_5$ and $x_i\ra -x_i, i=1,\ldots,4$.
It would be interesting to study the Kaluza-Klein spectrum in this case and compare 
it to the one found above. An analogy with the Type II case~\cite{ABKS} suggests that 
chiral operators coming from the ``Pfaffian'' worldsheet primary 
$({\bf N/2},{\bf N/2})$ have a nonperturbative origin: they are related to an 
M2-brane wrapped around submanifolds of ${\bf S}^4/({\bf Z}_2\times {\bf Z}_2)$. 
The fact that one of these operators is in the adjoint of $E_8\times E_8$ suggests
that the corresponding submanifold has boundaries~\cite{horwitt}.

{\it The E-series}

The E-series modular invariants have left-right symmetric primaries with the
following spins:
\begin{eqnarray}
&& E_6:\quad 2j=0,3,4,6,7,10\nn\\
&& E_7:\quad 2j=0,4,6,8,10,12,16\\
&& E_8:\quad 2j=0,6,10,12,16,18,22,28\nn
\end{eqnarray}
A primary with spin $j$ yields, among others, a chiral operator which
is a singlet of $G$ and transforms with respect to $SO(4)$ as a traceless symmetric 
tensor of order $2j+2$. It is easy to see that the orders of these tensors match 
the orders of the Casimirs of the respective Lie algebras.

\section{Fivebranes near orbifold singularities}

In this section we construct heterotic LSTs whose global symmetry is a proper
subgroup of $SU(2)_L\times SU(2)_R\times G$. The simplest way to obtain such
theories is to orbifold the ADE LSTs discussed above by a finite group
$\Gamma\subset SU(2)_L\times G$ (if one wants to preserve $(1,0)$ SUSY, $SU(2)_R$
has to be a spectator.) Since $SU(2)_L$ acts on the fundamental WZW field $g(z,\bar z)$
by left  multiplication, one can interpret the orbifolded LSTs as representing
fivebranes near an ${\bf R}^4/\Gamma$ orbifold. However, this interpretation can be
taken literally only when the dilaton gradient in the theory is small, i.e. 
for large $N$. In particular, orbifolds of $E_n$ LSTs cannot be understood in
geometric terms. Furthermore, for special values of the WZW level the worldsheet CFT 
has accidental symmetries, in which case $\Gamma$ can be chosen to be a subgroup
of this larger symmetry group.\footnote{This remark applies equally well to
Type II LSTs with $(1,0)$ SUSY.} For example, $SU(2)_4$ WZW corresponding to the 
modular invariant of type $D_4$ has accidental $SU(3)_1$ symmetry~\cite{bounahm}. 
In this paper we limit ourselves to studying ${\bf Z}_n$ orbifolds of the A-series,
leaving the investigation of ``nongeometric'' and nonabelian orbifolds for the future. 

Let $\gamma$ be a generator of ${\bf Z}_n$. Its acts on the WZW field $g(z,\bar z)$
via 
\be
g(z,\bar z)\ra e^{-\frac{2\pi i}{n}\sigma_3} g(z,\bar z).
\ee
The right-moving WZW currents are invariant, while the left-moving currents
transform as
\be
J_L(\bar z)\ra e^{-\frac{2\pi i}{n}\sigma_3} J_L(\bar z) e^{\frac{2\pi i}{n}\sigma_3}.
\ee
Thus the left-moving current algebra is broken down to $U(1)$ for $n>2$ and
left intact for $n=2$. 

We must also choose the action of $\gamma$ on the left-moving internal 
bosons $H_\alpha, \alpha=1,\ldots,16$. Without loss of generality we can take 
$\gamma$ which acts by a shift,
\be
H_\alpha\ra H_\alpha+2\pi v_\alpha.
\ee
$v$ is defined modulo vectors from ${\cal L}$ where ${\cal L}$ is the lattice on 
which the $H_\alpha$ live. In addition, the vector $v$ must satisfy
$nv\in {\cal L}$. 
We will define $u=nv$; possible choices of $u$ (and therefore $v$) are classified by 
${\cal L}/n{\cal L}$. Geometrically, picking $u$ amounts to picking a monodromy of the 
gauge bundle on ${\bf R}^4/{\bf Z}_n\backslash \{0\}$.

It is convenient to think about the monodromy as due to a point-like instanton sitting
at the origin of ${\bf R}^4/{\bf Z}_n$. The instanton charge depends on the monodromy
and can be determined by blowing up the orbifold. Then, as a consequence of the anomalous
Bianchi identity $dH=\Tr\ R\wedge R - \Tr\ F\wedge F$, the orbifold plane has a
fivebrane charge which depends on $u$. 

From the worldsheet point of view, it is more convenient to fix the WZW level
$k$ (the analog of the fivebrane charge) and ask what choices of $u$ are
allowed by modular invariance.
Below we show that modular invariance requires
\be\label{modular}
k+\frac{u^2}{2}=0\ mod\ n .
\ee
Note that since ${\cal L}$ is an even lattice, $u^2$ is always even, and therefore there 
is no conflict between modular invariance and integrality of $k$.

In order to implement the ${\bf Z}_n$ action on the the WZW model, it is
convenient to rewrite the left moving sector as a product of a parafermionic
$SU(2)/U(1)$ coset and a free scalar field $\xi$. The orbifold group acts only on the 
free scalar. For details see e.g.~\cite{KLL}.  The left-moving boson $\xi$ is
related to the current via $$J^3_L=\frac{ik}{2}\partial\xi.$$
$\xi$ has a unit radius and obeys an OPE of the form 
$$\xi(\bar z)\xi(0)= -\frac{2}{k} \log {\bar z} + \cdots.$$
$\gamma$ acts on $\xi$ via $$\xi\ra \xi+ \frac{4\pi}{n}.$$
The partition function of the orbifolded theory can be written schematically as
\be
Z_{orb} = \frac{1}{n} \sum_{p,q=0}^{n-1} (\gamma^p,\gamma^q),
\ee
where $(\gamma^p,\gamma^q)$ corresponds
to the contribution twisted by $\gamma^p$ with $\gamma^q$ inserted in the trace.
The untwisted partition function projected onto the invariant states is given by
the terms with $p=0$. The twist operator which
takes us to the sector twisted by $\gamma^w$ is given by
\be
\Omega_w\sim e^{\frac{iw}{n}(k\xi+u\cdot H)}
\ee
A primary from the $w^{\rm th}$ twisted sector has a generic form
\be
\Omega_w(\bar z) V_{j_R,m_R}(z) V_{j_L,m_L}(\bar z) e^{i\alpha\cdot H}.
\ee
where $\alpha\in {\cal L}$. (For $\alpha=0$ the last factor must be replaced 
with $\partial H$.) Requiring $h_L-h_R\in {\bf Z}$ we get
\be\label{levelm}
w\left(2m_L+\frac{wk}{n}\right)+w\alpha\cdot u +\frac{w^2 u^2}{2n}=0\ mod\ n.
\ee
Setting $w=1$ and taking into account that $2m_L\in {\bf Z}$ we obtain Eq.~(\ref{modular}).
Let us define an integer $\ell$ by $\ell=(k+u^2/2)/n$. Requiring that operators
in different sectors be mutually local we obtain a quantization condition on
$m_L$ in the $w^{\rm th}$ twisted sector:
\be
2m_L+\alpha\cdot u+w\ell = 0\ mod\ n
\ee
With this quantization condition Eq.~(\ref{levelm}) is satisfied for all $w$. 

We now consider some special choices for $u$. The simplest choice is $u=0$.
Then the action of $\gamma$ on $H_\alpha$ is trivial and the worldsheet CFT has
unbroken $G$ current algebra. A closely related CFT was considered in Ref.~\cite{KLL}.
From the space-time point of view $u=0$ means that the monodromy of the flat
connection on ${\bf R}^4/{\bf Z}_n\backslash\{0\}$ is taken to be trivial.
For $n>2$ we thus find an LST with global symmetry $U(1)_L\times SU(2)_R\times G$.
For $n=2$ $\gamma$ acts trivially on the $SU(2)_L$ currents and the symmetry is
$SU(2)_L\times SU(2)_R\times G$. The $n=2$ model is identical to the D-series LSTs
of Section II. 

For $G=Spin(32)/{\bf Z}_2$ we may also take $u=(2,0,\ldots,0).$ For $n>2$ the monodromy 
corresponding to this $u$ breaks the left-moving internal current algebra from $so(32)$ down to 
$so(30)\oplus u(1)$. The case $n=2$ is special in that
none of the gauge bosons of $G$ are projected out and the gauge symmetry
algebra is still $so(32)$ and the $su(2)_L$ also remains unbroken. Thus the
symmetry algebra is the same as for
the D-series LSTs. Moreover, for $n=2$ the condition 
Eq.~(\ref{modular}) becomes $k+2=0\ mod\ 2$, which is the same as for the D-series LSTs.
Nonetheless, this theory is not identical to the D-series LST, because $\gamma$ acts
nontrivially on spinors of $Spin(32)/{\bf Z}_2$. As a result, the global symmetry group of
this theory is $(Spin(32)/{\bf Z}_2\times SU(2)_L)/{\bf Z}_2$, where the second ${\bf Z}_2$
reverses the sign of the spinor of $Spin(32)/{\bf Z}_2$ and generates the center of $SU(2)_L$.

For $G=E_8\times E_8$ an analogous choice is $u=(2,0,\ldots,0)$ for the
first $E_8$ and $u=(0,\ldots,0)$ for the second $E_8$. For $n>2$ this leads to
LSTs with global symmetry algebra $u(1)_L\oplus u(1)\oplus so(14)\oplus e_8$.
For $n=2$ we get instead $su(2)_L\oplus so(16)\oplus e_8$.

None of the above models corresponds to fivebranes near a perturbative
heterotic orbifold with the standard embedding of the spin connection into the gauge
connection. The latter has a monodromy which breaks $so(32)$ down to $su(2)\oplus u(1)\oplus
so(28)$ and $e_8\oplus e_8$ down to $u(1)\oplus e_7\oplus e_8$. (For $n=2$ it only
breaks $so(32)$ down to $su(2)\oplus su(2) \oplus so(28)$ and 
$e_8\oplus e_8$ down to $su(2)\oplus e_7\oplus e_8$.) To get such a spectrum of unbroken
gauge bosons we need to set $u=(1,1,0,\ldots,0)$ for $G=Spin(32)/{\bf Z}_2$;
for $G=E_8\times E_8$ we need to set $u=(1,1,0,\ldots,0)$ for the first $E_8$
and $u=(0,\ldots,0)$ for the second $E_8$.

\section{Conclusions}

In this paper we constructed a variety of heterotic LSTs arising from heterotic
string theory in linear dilaton backgrounds. For each affine Dynkin diagram
of type A, D, or E we constructed a worldsheet CFT which is holographic to a
heterotic $(1,0)$ LST with global symmetry $SU(2)_L\times SU(2)_R\times G$, 
where $SU(2)_R$ is the R-symmetry and $G=Spin(32)/{\bf Z}_2$ or $E_8\times E_8$.
The A-series are heterotic fivebranes in flat space. The D-series correspond to 
fivebranes near an ${\bf R}^4/{\bf Z}_2$ orbifold with a trivial gauge connection. 
The E-series do not admit a geometric description either in heterotic string theory or 
M-theory. In particular, we found that a single heterotic fivebrane admits a
well-defined holographic dual (for either choice of $G$). This is in contrast
with a single Type II fivebrane, which apparently does not lead to a decoupled
Poincare-invariant theory and does not admit a holographic description~\cite{ds}.

We studied the spectrum of chiral operators in these theories using worldsheet
methods. We found that in all models there are operators which are
singlets of $G$ and are traceless symmetric tensors of $SO(4)\sim SU(2)_L\times SU(2)_R$.
The orders of the tensors are in one-to-one correspondence with the orders of
the Casimirs of the corresponding ADE Lie group. This part of the spectrum is
very similar to that of Type II LSTs~\cite{ABKS}. There are also operators 
which are charged under $G$. However, comparison with the large $N$ limit
revealed that many chiral operators are missed by the worldsheet analysis. In
particular, we did not find any self-dual three-forms expected for
$E_8\times E_8$ LSTs, or any evidence for the $Sp(k+1)$ gauge group for $\spin$
LSTs. 

We compared the chiral operators of the LST describing $\spin$ heterotic
fivebranes with chiral operators of the Yang-Mills theory describing Type I D5
branes. We found that the adjoints of $\spin$ match precisely between the two
theories. The above-mentioned traceless symmetric tensors of $SO(4)$ also
almost match, but for $N_5>1$ the LST has one more of them than the Yang-Mills theory. 
We suggested several possible viewpoints on this discrepancy. The most intriguing
one is that along some flat directions of the LST certain modes become
free in the Yang-Mills theory but not in the full LST. If this possibility is
realized, then the linear dilaton backgrounds we considered describe heterotic
LSTs infinitely far along such flat directions rather than at the origin of the
moduli space. 

Finally, we have constructed linear dilaton backgrounds which describe heterotic
fivebranes near orbifold singularities of the form ${\bf R}^4/{\bf Z}_n$. 
The orbifold does not have to be a perturbative orbifold with zero fivebrane
charge; rather, when the fivebrane charge of the orbifold plane is nonzero, modular
invariance requires the level of the WZW model to be shifted appropriately 
(see Eq.~(\ref{modular})).
We also pointed out that it may be possible to construct nongeometric backgrounds with 
linear dilaton by gauging accidental symmetries of WZW models at special values of 
the level. It would be interesting to check whether modular-invariant orbifolds of this 
sort really exist.

\acknowledgements{We would like to thank Micha Berkooz, 
Emanuel Diaconescu, Nati Seiberg, and Angel Uranga for helpful discussions.
The work of M.G~was supported in part by DOE grants \#DF-FC02-94ER40818 and
\#DE-FC-02-91ER40671, while
that of A.K. by DOE grant \#DE-FG02-90ER40542.}

{\tighten

}

\end{document}